\documentclass[a4paper,11pt]{article}
\usepackage{pos}
\graphicspath{{Figures/}{./}} 

\newcommand{\beq}{\begin{eqnarray}}
\newcommand{\eeq}{\end{eqnarray}}

\newcommand{\npo}{{n+1}}
\newcommand{\npt}{{n+2}}

\newcommand{\one}{\, (\mathbf{1})}
\newcommand{\two}{\, (\mathbf{2})}
\newcommand{\otwo}{\, (\mathbf{12})}

\newcommand{\RV}{(\mathbf{RV})}

\newcommand{\nnb}{\nonumber}

\title{Analysis of (n+1) and n-parton contributions for computing QCD jet cross sections in the local analytic subtraction scheme}
\ShortTitle{RV and VV in the LASS}

\author*[a]{Bakar Chargeishvili}
\author[b]{Giuseppe Bevilacqua}
\author[c,d]{Adam Kardos}
\author[a]{Sven-Olaf Moch}
\author[d]{Zoltán Trócsányi}

\affiliation[a]{Universität Hamburg, II. Institute for Theoretical Physics,\\
Luruper Chaussee 149, 22761 Hamburg, Germany}

\affiliation[b]{Institute of Nuclear and Particle Physics, NCSR "Demokritos",
	15341 Agia Paraskevi, Greece
}
\affiliation[c]{University of Debrecen, Faculty of Science and Technology, Department of Experimental Physics, 4010,\\
Debrecen, PO Box 105, Hungary}
\affiliation[d]{Institute for Theoretical Physics, ELTE Eötvös Loránd University, Pázmány Péter sétány 1/A,\\
H-1117 Budapest, Hungary}

\emailAdd{bakar.chargeishvili@desy.de}
\emailAdd{bevilacqua@inp.demokritos.gr}
\emailAdd{kardos.adam@science.unideb.hu}
\emailAdd{sven-olaf.moch@desy.de}
\emailAdd{zoltan.trocsanyi@cern.ch}

\abstract{
We analyze and implement the Local Analytic Sector Subtraction (LASS) scheme
for handling infrared singularities in next-to-next-to-leading order (NNLO)
calculations in perturbative QCD. We examine the key aspects of the scheme
including sector function construction, singular limit parametrization,
subtraction counterterm derivation, and integration techniques. As a
proof-of-concept, we numerically implement LASS for the process $e^+e^-
\rightarrow 3$ jets. In this study we examine the limiting behavior of
subtraction terms for real-virtual contribution and explicitly demonstrate the
pole cancellation of the double-virtual contribution.
Differential cross sections are computed for several event shape observables,
showing the stability and efficiency of LASS scheme. This work lays the
foundation for developing an automated tool for NNLO QCD calculations using
this promising scheme.
}

\FullConference{Loops and Legs in Quantum Field Theory (LL2024)\\
 14-19, April, 2024\\
Wittenberg, Germany\\}

\begin{document}
\emergencystretch 3em
\maketitle

\section{Introduction}
Precise theoretical predictions are crucial for interpreting collider data and testing the Standard Model.
Next-to-next-to-leading order (NNLO) accuracy in QCD is now needed for many key processes.
However, at NNLO, handling the infrared divergences from soft and
collinear radiation is very challenging.
Several approaches have been developed to treat these singularities at NNLO,
including slicing methods~\cite{Frixione:2004is,Anastasiou:2003gr}, non-local
subtraction schemes~\cite{Boughezal:2015aha,Gaunt:2015pea}, and local
subtraction methods like antenna
subtraction~\cite{Gehrmann-DeRidder:2005btv,Currie:2013vh},
CoLoRFulNNLO~\cite{Somogyi:2005xz,Somogyi:2006da,Somogyi:2006db}, nested
soft-collinear subtraction~\cite{Caola:2017dug} and the recently proposed Local
Analytic Sector Subtraction (LASS)~\cite{LASS1,LASS_INT,LASS2}.
LASS aims to provide a fully local and analytic subtraction framework that can
be implemented numerically within any existing code, using sector decomposition
of the phase space and flexible parametrizations of singular limits to derive
counterterms that can be integrated analytically.
In this work, we implement LASS and apply it to the process $e^+e^-
\rightarrow$ 3 jets as a proof-of-concept. We analyze the theoretical framework
in depth, perform rigorous checks of the counterterms, and compute differential
distributions for event shape observables.

\section{Local Analytic Sector Subtraction at NNLO}
The NNLO correction to a cross section schematically takes the form:
\begin{align}
d\sigma^\mathrm{NNLO} = \int d\Phi_n VV \delta_n(X) + \int d\Phi_{n+1} RV \delta_{n+1}(X)
\int d\Phi_{n+2} RR \delta_{n+2}(X)
\end{align}
Here $VV$, $RV$ and $RR$ denote the double-virtual, real-virtual and
double-real squared matrix elements, and $\delta_n(X)$ fixes the observable $X$
at its value for $n$-body kinematics.
To numerically integrate each term, we must either eliminate or regulate
all singularities. LASS achieves this by adding and subtracting
counterterms: [replace sub with reg for regularized]
\beq
\label{eq:NNLO_RR_sub}
RR_{\, \rm sub}(X)
& \equiv &
RR \, \delta_\npt (X)
-
K^{\one} \, \delta_\npo(X)
-
\left( K^{\two} - K^{\otwo} \right) \delta_n (X)
\, ,
 \\
\label{eq:NNLO_RV_sub}
RV_{\, \rm sub}(X)
& \equiv &
\left(RV + I^{\one} \right) \delta_\npo(X)
\, - \,
\left( K^{\RV} + I^{\otwo} \right) \delta_n(X)
\, ,
\\
\label{eq:NNLO_VV_sub}
VV_{\, \rm sub}(X)
& \equiv &
\left( VV + I^{\two} + I^{\RV} \right) \delta_n (X)
\, .
\eeq
$K^{\one}$ is a counterterm that accounts for the single unresolved
singularities, $K^{\two}$ accounts for the double unresolved singularities, and
$K^{\otwo}$ is introduced to avoid double counting by removing the overlap
between $K^{\one}$ and $K^{\two}$. At RV level there is still a possibility to
introduce an additional subtraction term $K^{\RV}$ which deals with the
phase-space singularities of RV contribution. So in total one needs 4 kinds of
subtraction terms. Then those subtraction terms are integrated
analytically over the factorized singular regions of the radiation
phase spaces and the integrated versions of those subtraction terms are then
added back. They are defined as:
\beq
\label{eq:intcountNNLO}
  I^{\one} \, \equiv \, \int d \Phi_{{\rm rad}} \, K^{\one} \, ,
  && \qquad
  I^{\two} \, \equiv \, \int d \Phi_{{\rm rad}, 2} \, K^{\two} \, ,
  \nnb \\
  I^{\otwo} \, \equiv \, \int d \Phi_{{\rm rad}} \, K^{\otwo} \, ,
  && \qquad
  I^{\RV} \, \equiv \, \int d \Phi_{{\rm rad}} \, K^{\, \RV} \, ,
\eeq
where $\Phi_{{\rm rad}}$ parametrizes the phase-space of the single unresolved
radiation and $\Phi_{{\rm rad},2}$ - the phase-space of the double unresolved
radiation.
To construct the counterterms, LASS partitions the phase space into
sectors containing a minimal set of singularities using sector functions
$W_{abcd}$ defined in Refs.~\cite{LASS1,LASS2}, see also Ref.~\cite{Kardos} in
these proceedings.

Various singular projectors $S_i$, $C_{ij}$ etc. are defined to parametrize different unresolved configurations. The counterterms are built by applying these projectors to the matrix element in each sector.
Integrating the counterterms over the unresolved phase space is a key aspect of
LASS. Flexible mappings of the phase space are used to factorize the integrals
into a radiation part and a Born-level part. To this end the Catani-Seymour
final-state mapping~\cite{Catani:1996vz} is being used.
This allows the counterterms to be analytically integrated over the radiation
variables, yielding expressions that can be added back to the singular
contributions in Eqs.~\eqref{eq:NNLO_RR_sub}-~\eqref{eq:NNLO_VV_sub}.

\section{Validation and results for $\mathbf{e^+e^- \rightarrow}$ 3 jets}
As a proof-of-concept, we implemented the LASS scheme for the process $e^+e^-
\rightarrow 3$ jets. At NNLO, this involves the subprocesses:
\begin{alignat}{3}
RR:&\qquad e^+e^- \to q \bar q q \bar q g \,, \qquad
&&e^+e^- \to q \bar q r \bar r g \,, \qquad
&&e^+e^- \to q \bar q g g g \label{eq:RRproc}\,,\\
RV:&\qquad e^+e^- \to q \bar q q \bar q \,, \qquad
&&e^+e^- \to q \bar q r \bar r \,, \qquad
&&e^+e^- \to q \bar q g g \label{eq:RVproc}\,,\\
VV:&\qquad e^+e^- \to q \bar q g\,, \label{eq:VVproc}
\end{alignat}
where the notation $q$ and $r$ is used to distinguish between quarks of
different flavors. The processes in Eq.~\eqref{eq:RRproc} involve only
tree-level kinematics and helicity amplitudes, from which the squared
matrix elements can be constructed using the results in
Refs.~\cite{Falck:1989uz,Hagiwara:1988pp,Berends:1988yn}. The processes
in Eq.~\eqref{eq:RVproc} contain one-loop virtual corrections, and
their matrix elements can be constructed using the method described in
Ref.~\cite{Bern:1993mq}. Finally, the process in Eq.~\eqref{eq:VVproc}
involves two-loop virtual corrections, and the corresponding matrix
element can be found in Refs.~\cite{Garland:2001tf,Gehrmann-DeRidder:2007vsv}.

The integration over physical phase space is managed using an in-house
implementation of the {\mbox VEGAS+} algorithm~\cite{Lepage:2020tgj}.

\subsection{Pole cancellation and limiting behavior}
As a first check, we verified the pole structure of the $VV$
contribution by comparing the $\epsilon$ expansion of
$I^{(2)}+I^{(RV)}$ with the known singularities of the two-loop virtual
correction. We found perfect agreement.

Next, we studied the limiting behavior of the $RV$ counterterms in all
singular regions of phase space.  Since $K^{\RV}$ is designed to cancel
the phase-space singularities of $RV$, and $I^{\otwo}$ cancels the
singularities of $I^{\one}$, it is sufficient to examine their
respective ratios in the singular limits appearing in the $K^{\RV}$
term. This has been done for each individual subprocess listed in
Eq.~\eqref{eq:RVproc}.  The $e^+e^- \to
q\bar{q}q\bar{q}$ process features a total of $4$ singular limits,
$e^+e^- \to q\bar{q}r\bar{r}$ features 2, and $e^+e^- \to q\bar{q}gg$
features $7$ singular limits. For illustration purposes, in this
proceedings we present plots obtained for the real virtual and
double virtual processes with four equal flavour quarks in the final
state. Our results for the other final states are very similar.

\begin{figure}[htbp]
    \centering
    \includegraphics[width=0.49\textwidth]{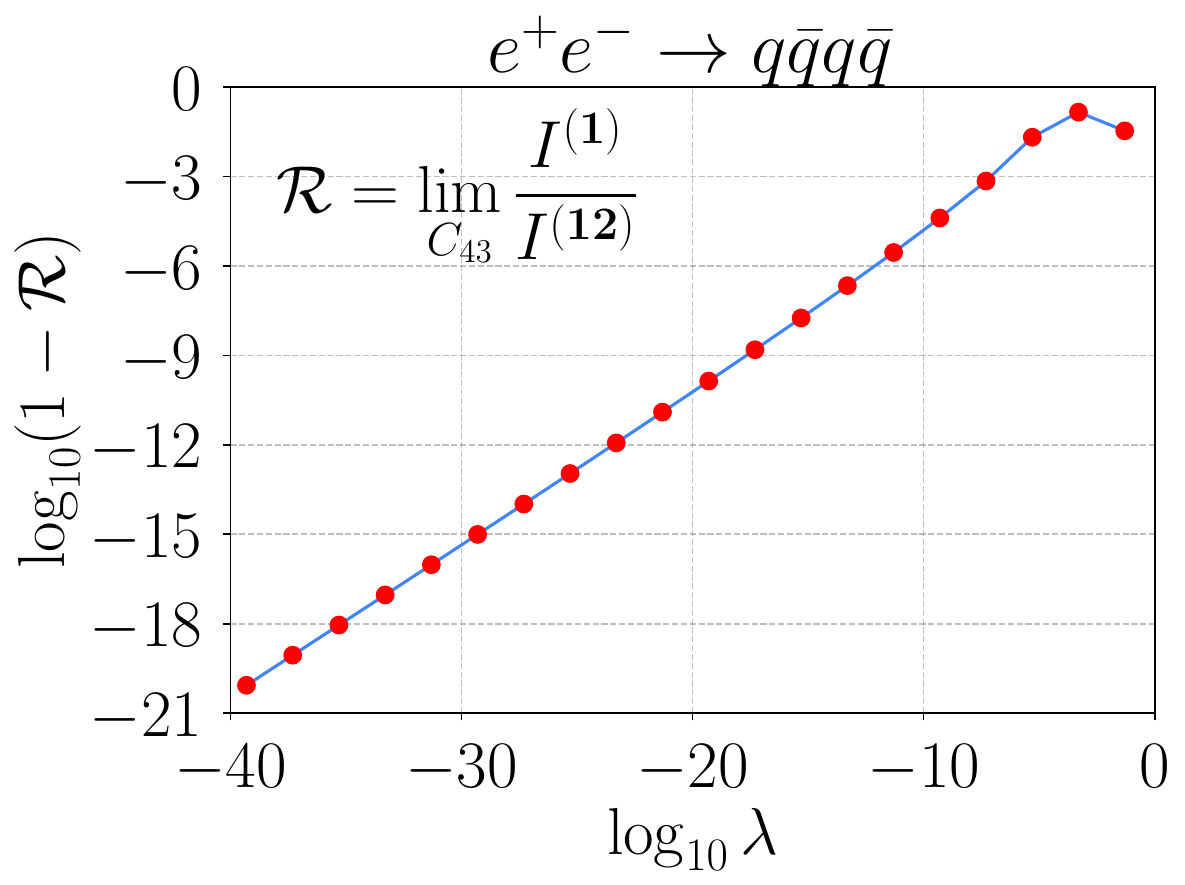} ~
    \includegraphics[width=0.49\textwidth]{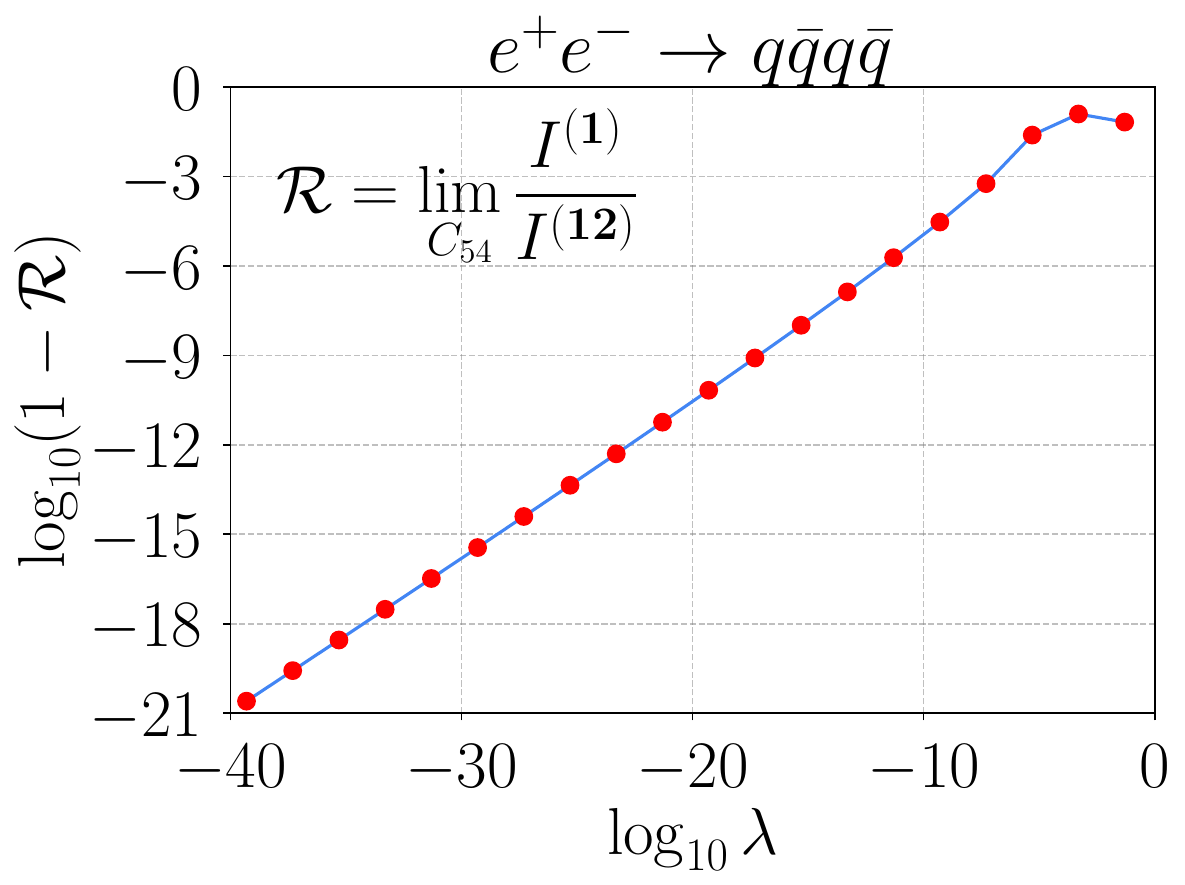}\\
    \includegraphics[width=0.49\textwidth]{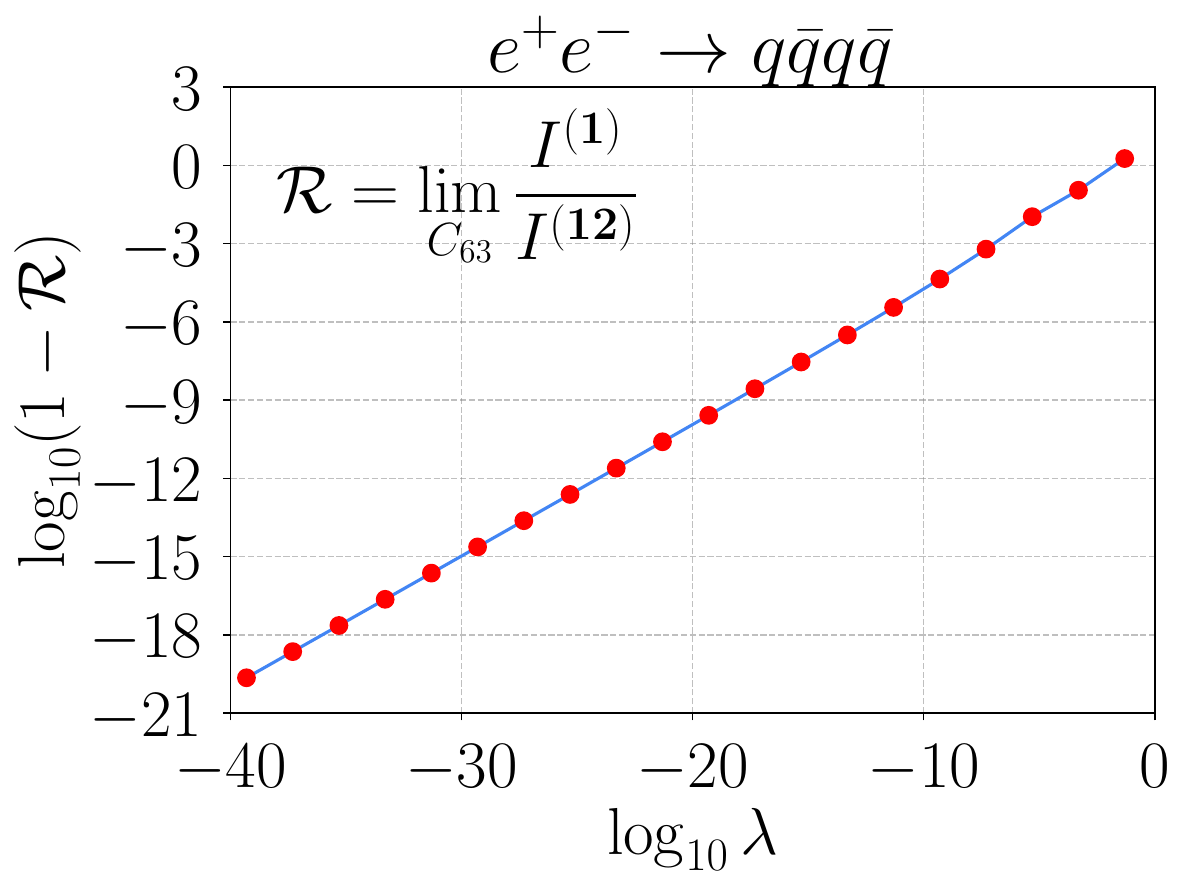} ~
    \includegraphics[width=0.49\textwidth]{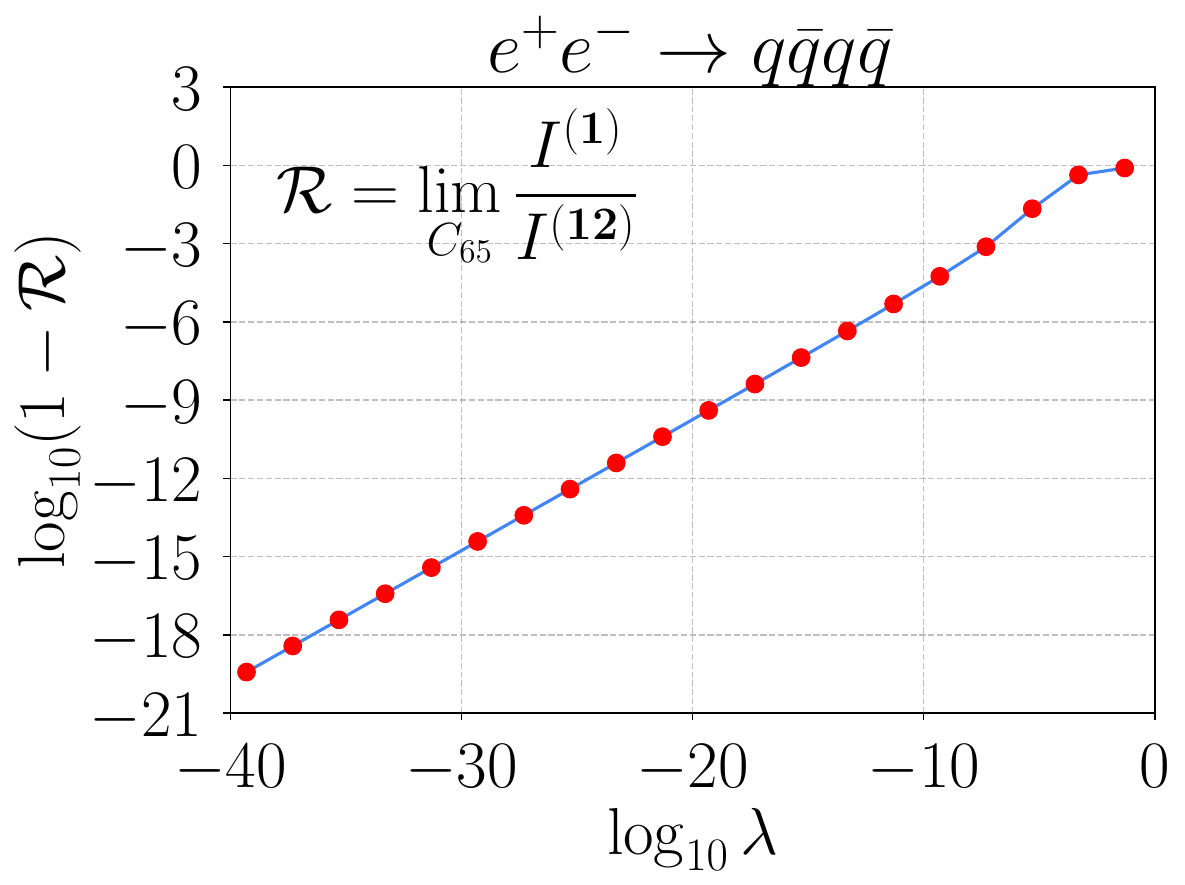}\\
    \includegraphics[width=0.49\textwidth]{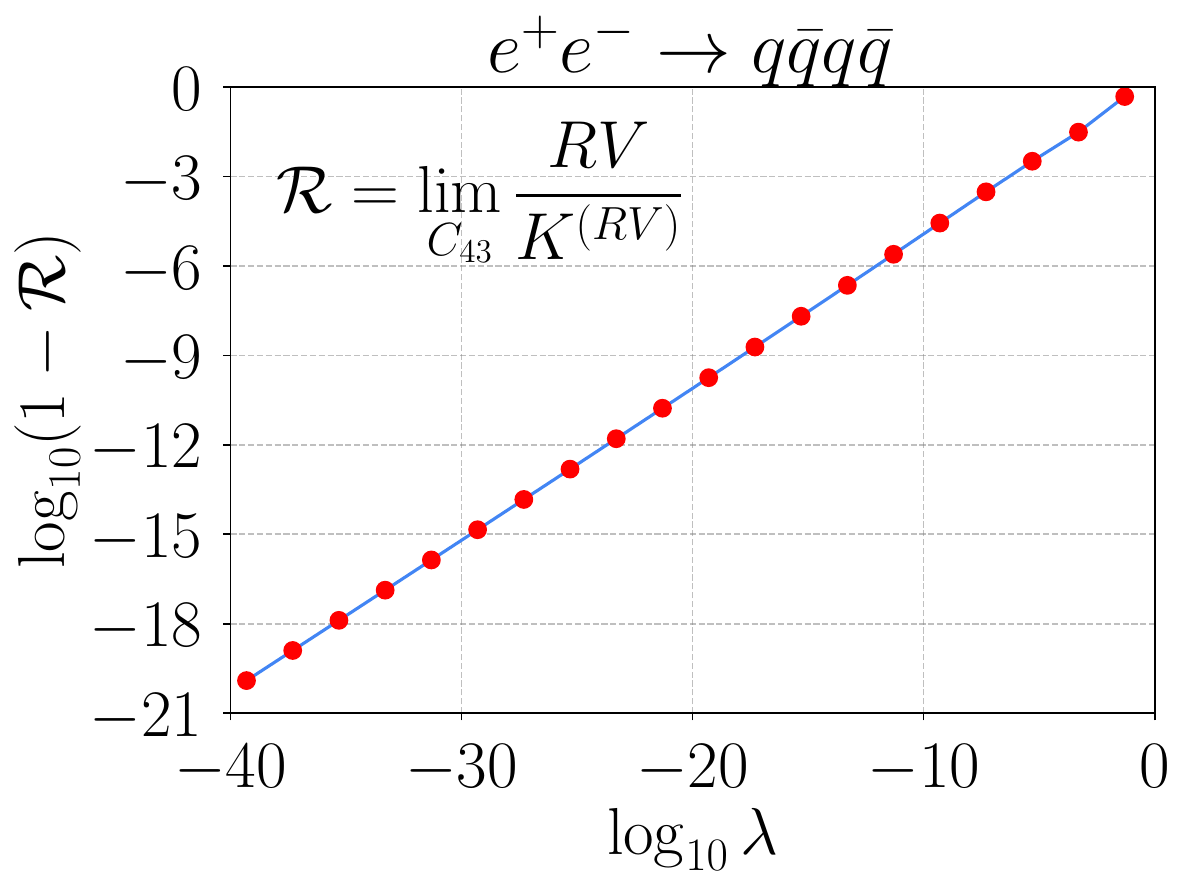} ~
    \includegraphics[width=0.49\textwidth]{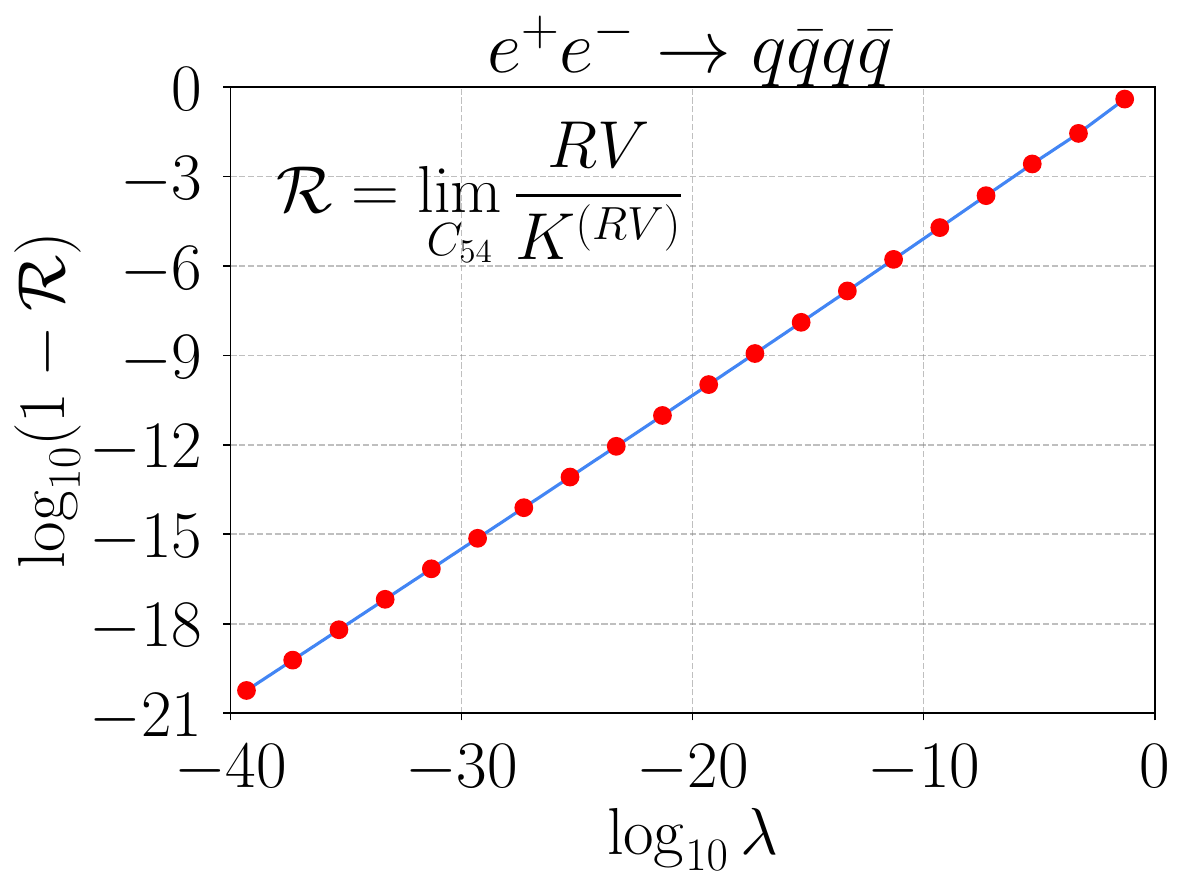}\\
    \includegraphics[width=0.49\textwidth]{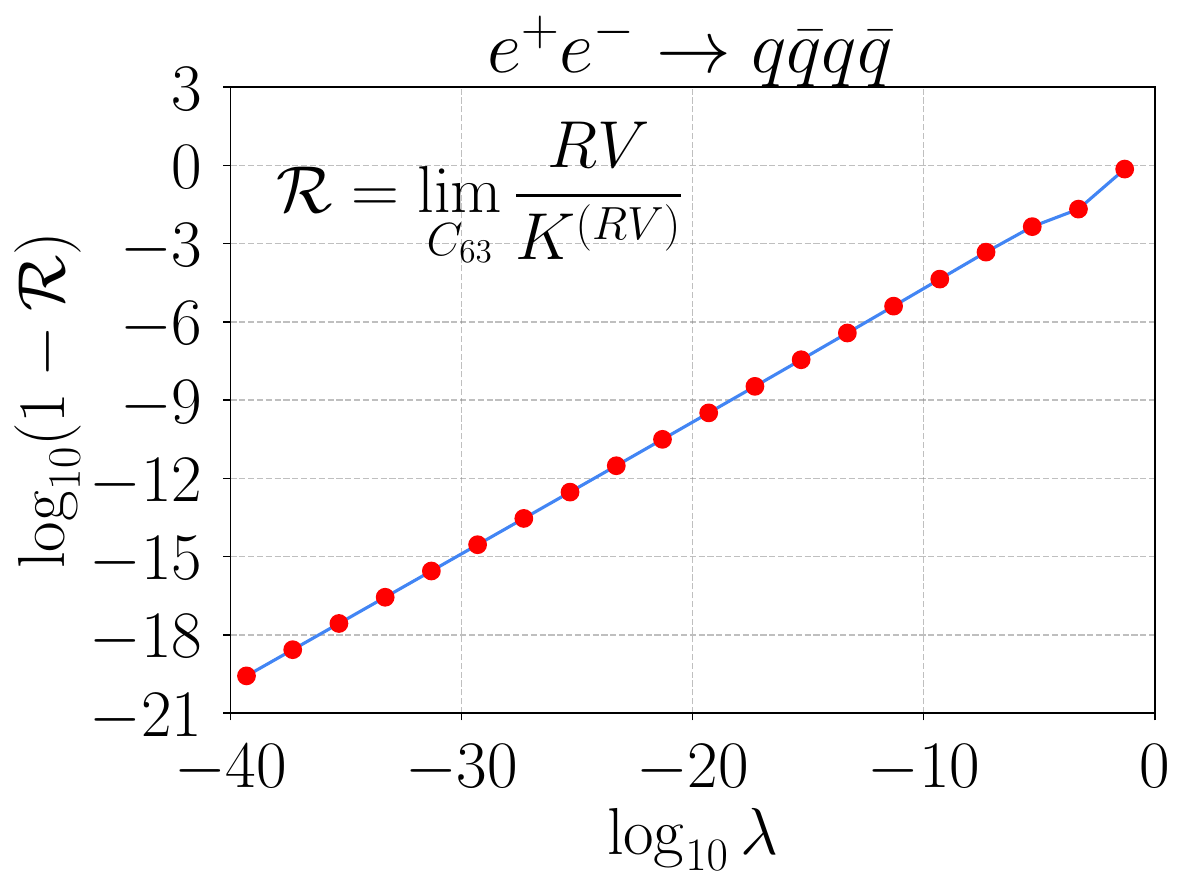} ~
    \includegraphics[width=0.49\textwidth]{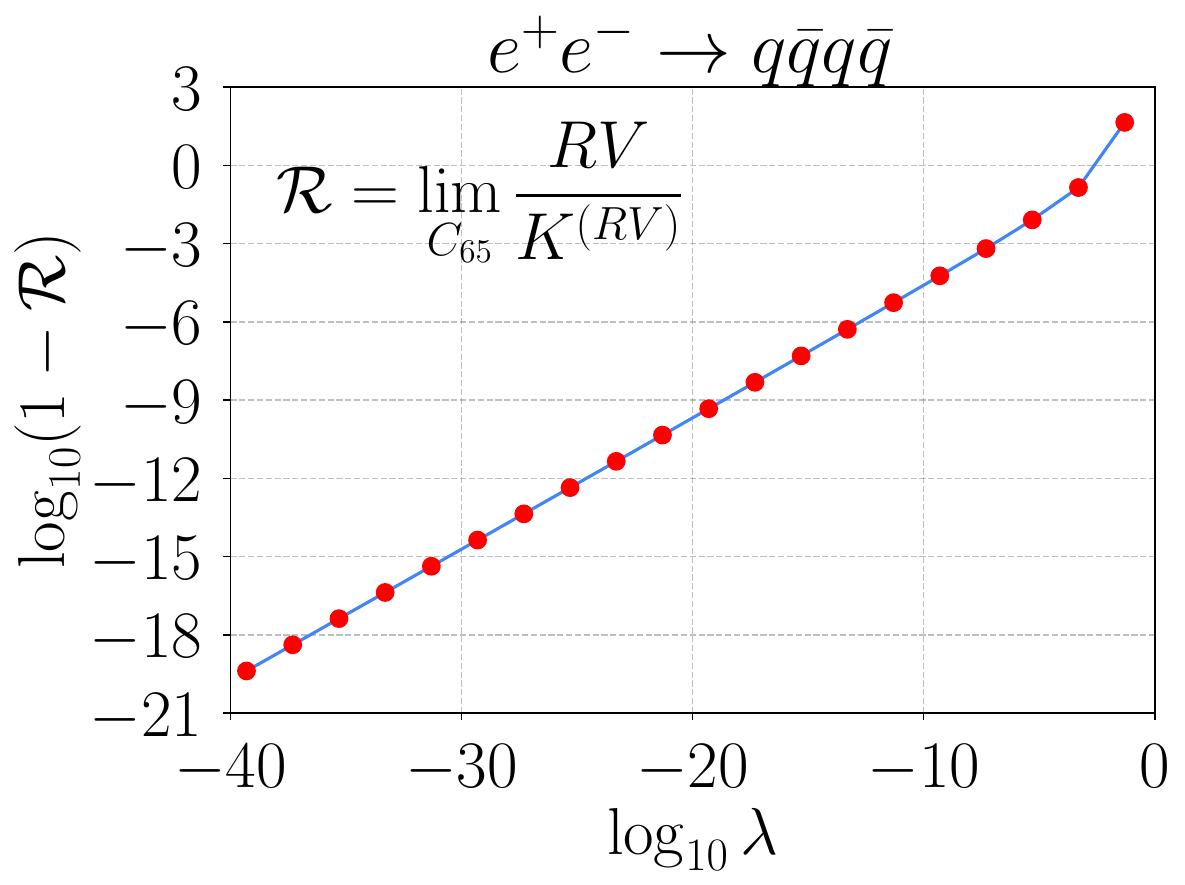}
\caption{
Limiting behavior of the ratio between $I^{\one}$ and $I^{\otwo}$ terms
(top row) and $RV$ and $K^{\RV}$ terms (bottom row) in various singular
limits for the one-loop correction to $e^+e^-\to q\bar{q}q \bar{q}$
contribution. Each plot corresponds to a different singular limit, as
indicated on the respective plot.
}
\label{fig:RV_limits_qq}
\end{figure}

The results for the ratios of $I^{\one}$ and $I^{\otwo}$ pair and $RV$
and $K^{\RV}$ pair for $e^+e^-\to q\bar{q}q \bar{q}$ contribution are
plotted in Fig.~\ref{fig:RV_limits_qq}. {The precise way of taking these
limits together with the definition of the parameter $\lambda$ that
controls the limit is described in the contribution \cite{Kardos} to
these proceedings.} On these plots, the absolute value of the numbers
read from the vertical axis indicates the number of digits of accuracy
with which the ratio matches unity and the notation
"$\lim\limits_{P}$" means that we are approaching that particular
limit. For example, "$\lim\limits_{C_{ijk}}$" corresponds to the case
when partons $i$, $j$, and $k$ become collinear.

\subsection{Technical cut and event shape distributions}
Making a step closer towards physical predictions, we computed
computed differential distributions for several event shape observables.
For this purpose, we choose four event shape observables typically studied in
$e^+e^-\to 3$ jet processes~\cite{DelDuca:2016csb, Weinzierl:2009ms}. The
observables of our choice are:
$\tau$-parameter (thrust),
$C$-parameter,
Energy-energy correlation,
Jet-cone energy fraction.
These observables are chosen because they provide complementary information
about the event topology and the dynamics of the final-state particles,
allowing for a comprehensive study of the underlying QCD processes.
In this study we use the definitions taken from~\cite{DelDuca:2016csb}.

When computing such distributions, due to the finite precision of
representing real numbers on the computer, one always have to introduce
a technical cut parameter $y_\mathrm{min}$ on the phase space to avoid
numerical instabilities in the deep IR regions.
Similarly to Ref.~\cite{Kardos:2019iwa} we are using the following cut to
exclude the unstable regions:
\begin{equation}
		y_{\min}=\underset{i, j}{\min}\frac{s_{ij}}{s}\, ,
		\label{eq:y_min}
\end{equation}

\begin{figure}[htb]
	\centering
	\includegraphics[width=0.49\textwidth]{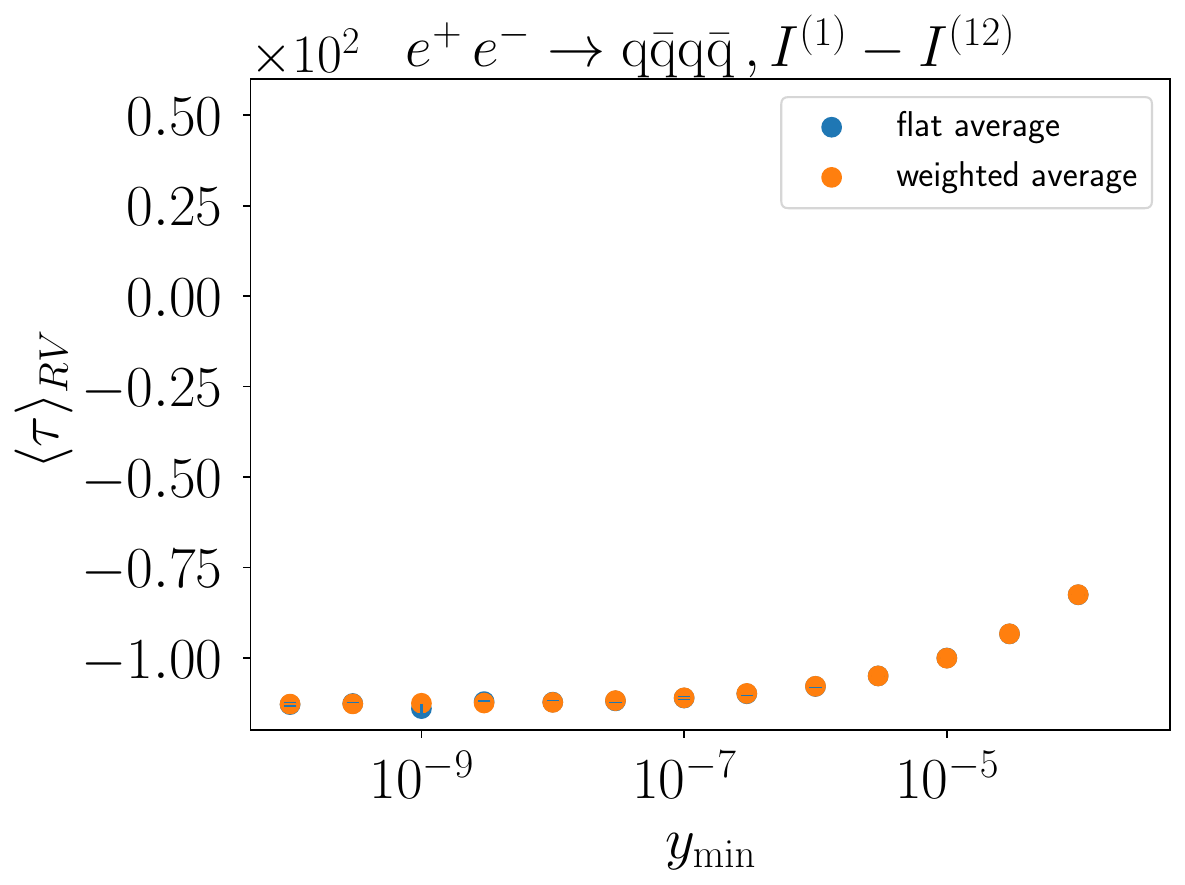}~
	\includegraphics[width=0.49\textwidth]{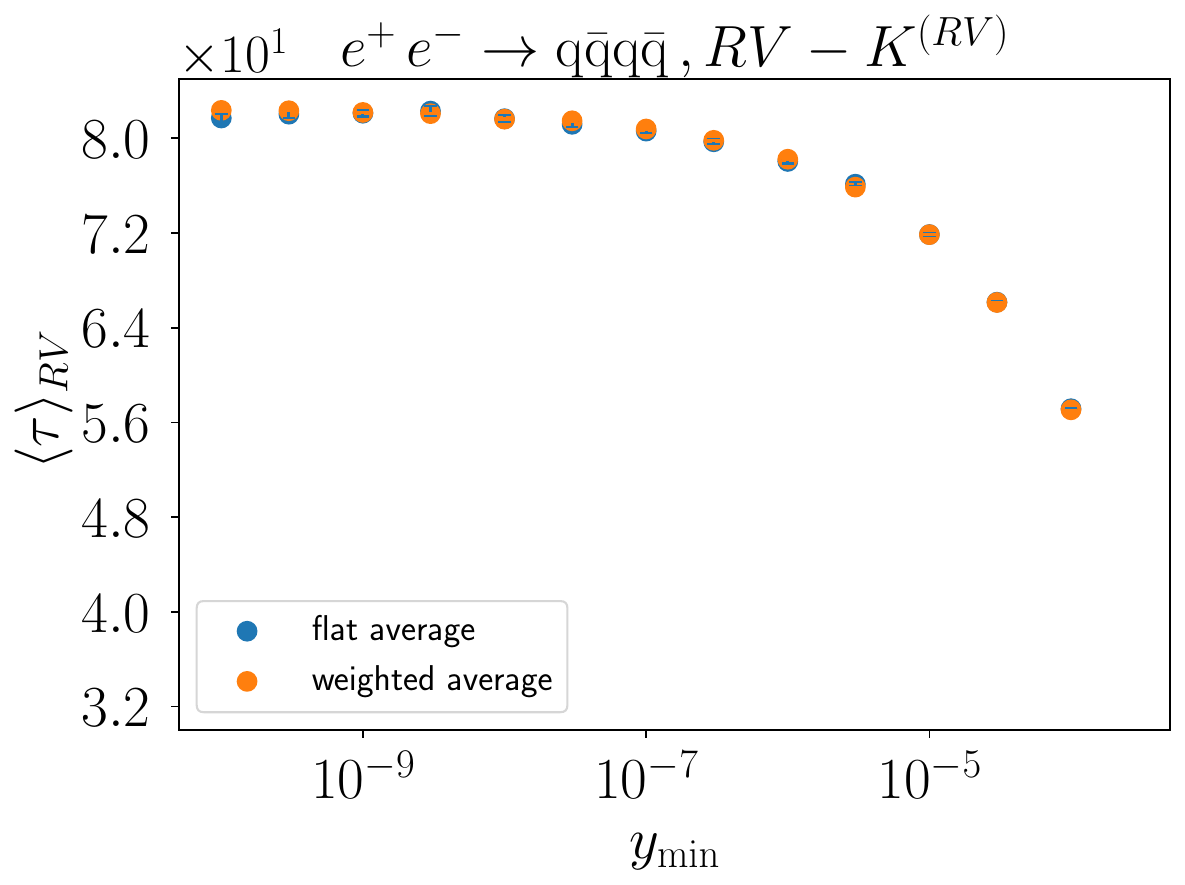}\\
	\caption{
		The saturation plots of $I^{\one}-I^{\otwo}$ (to the left) and
		$RV-K^{\RV}$ (to the right) contributions to the mean value of the
		thrust distribution by the process $e^+e^-\to q\bar q q \bar q$.
	}
	\label{fig:sat}
\end{figure}

To investigate the impact of this cut we perform the following study:
We set the initial value of $y_{\min}$ to $10^{-4}$ and gradually decrease it
in small intervals until it reaches to $10^{-10}$. For each value of
$y_{\min}$, we perform the Monte Carlo integration by generating
$10^7$
phase-space points and optimizing the grid $40$ times. After every iteration,
we calculate a new estimate of the cross-section. Once all $40$ iterations are
completed, we plot the first moment of the thrust distribution
as a function of the technical cut $\langle \tau \rangle(y_{\min})$.
To gain further insights, we combine the results of these $40$
independent estimates in each bin using two methods. First, we
calculate the average of the estimates. Second, we calculate the
weighted average, where the inverse values of the mean squared
uncertainties are used as weights. This latter
method effectively gives preference to more stable integration
iterations with smaller uncertainties.  The results of this study are depicted
in Fig.~\ref{fig:sat}.

We observe that in all studied cases a very good saturation is achieved at the
value $y_{\min} = 10^{-8}$. We also considered the different formulations of
$y_{\min}$ and for $RV$ contribution no significant effect was found.
Hence, for the subsequent studies we fix the value of $y_{\min}$ as defined
in Eq.~\eqref{eq:y_min} to $10^{-8}$.

\begin{figure}[htb]
	\centering
	\includegraphics[width=0.49\textwidth]{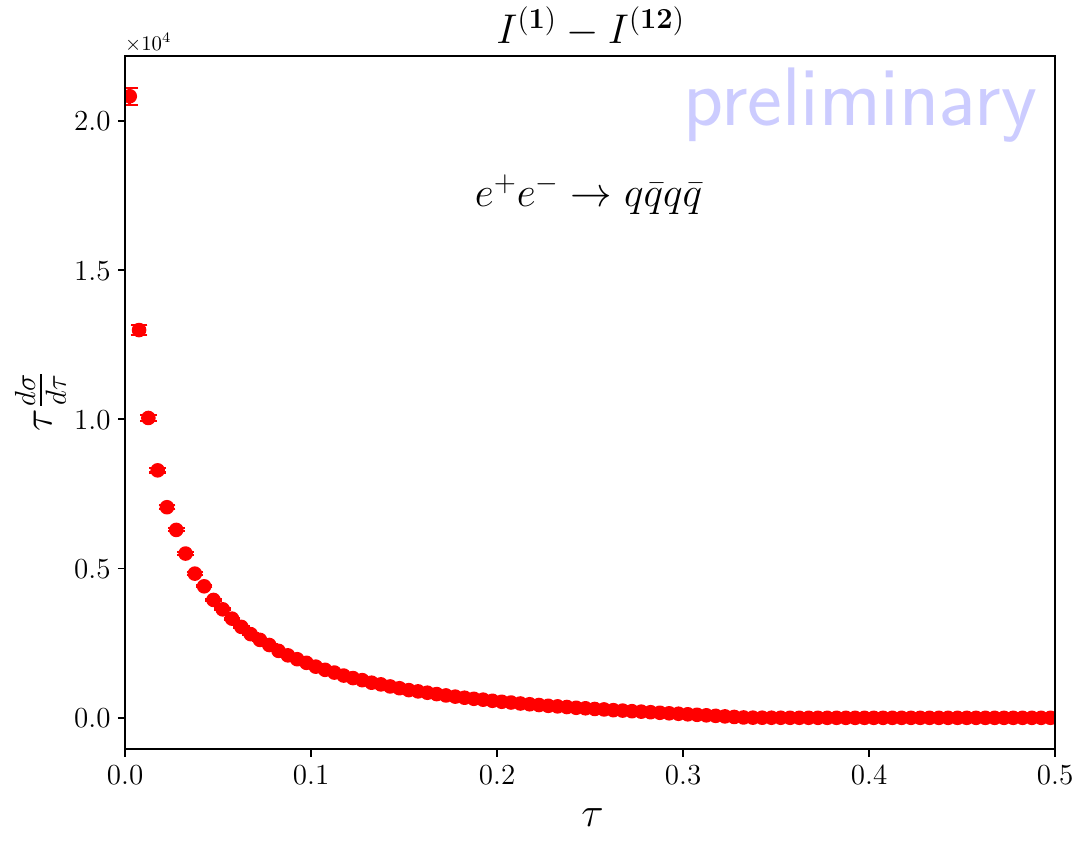}~
	\includegraphics[width=0.49\textwidth]{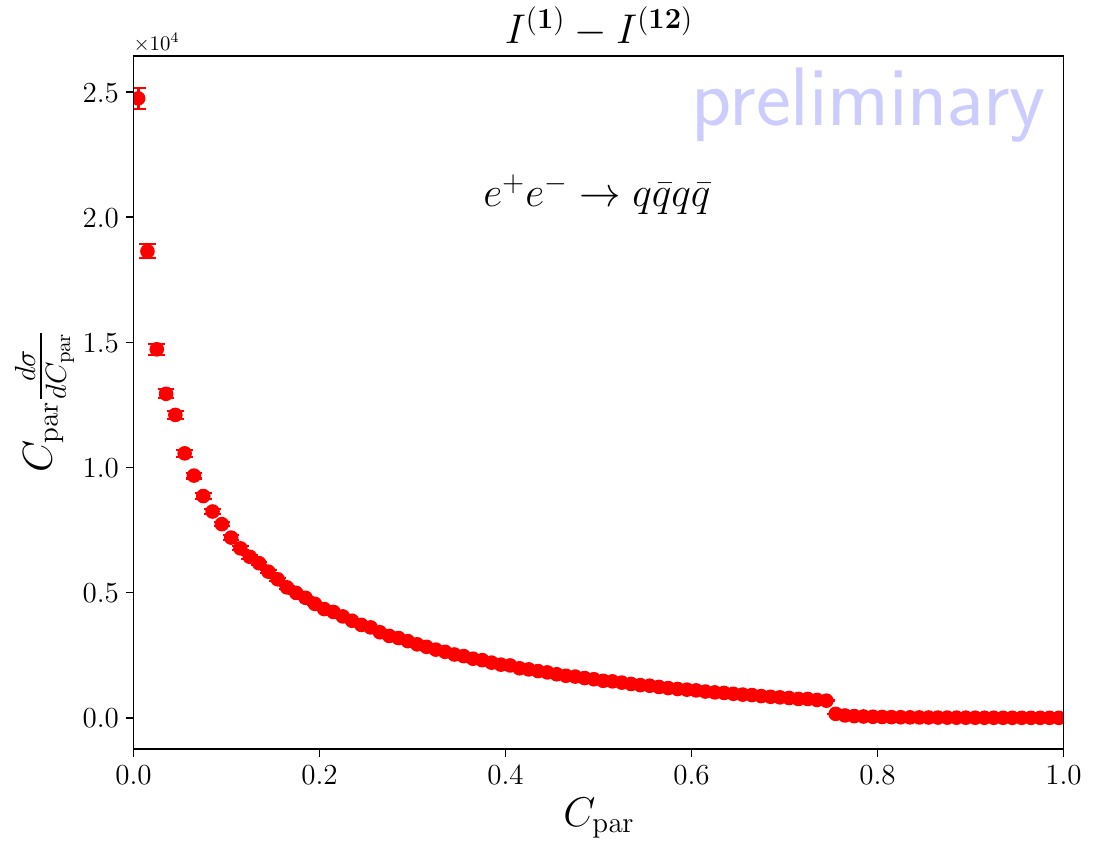}\\
	\includegraphics[width=0.49\textwidth]{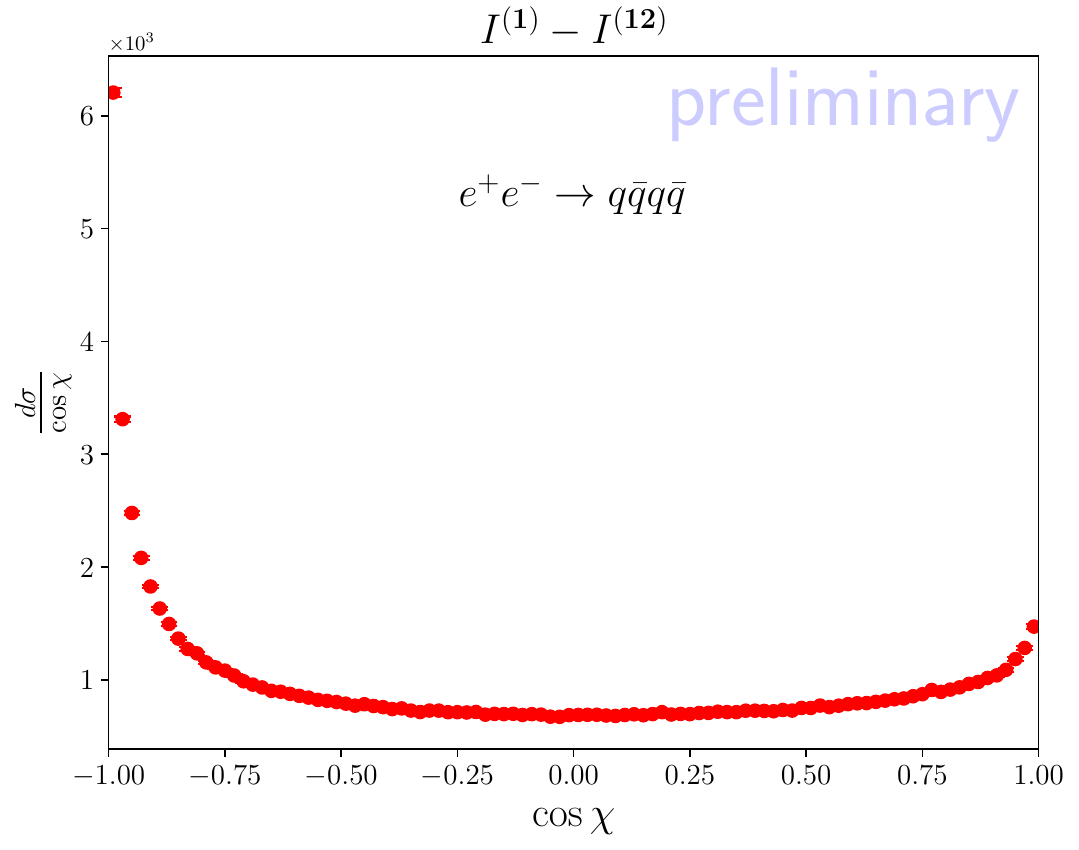}~
	\includegraphics[width=0.49\textwidth]{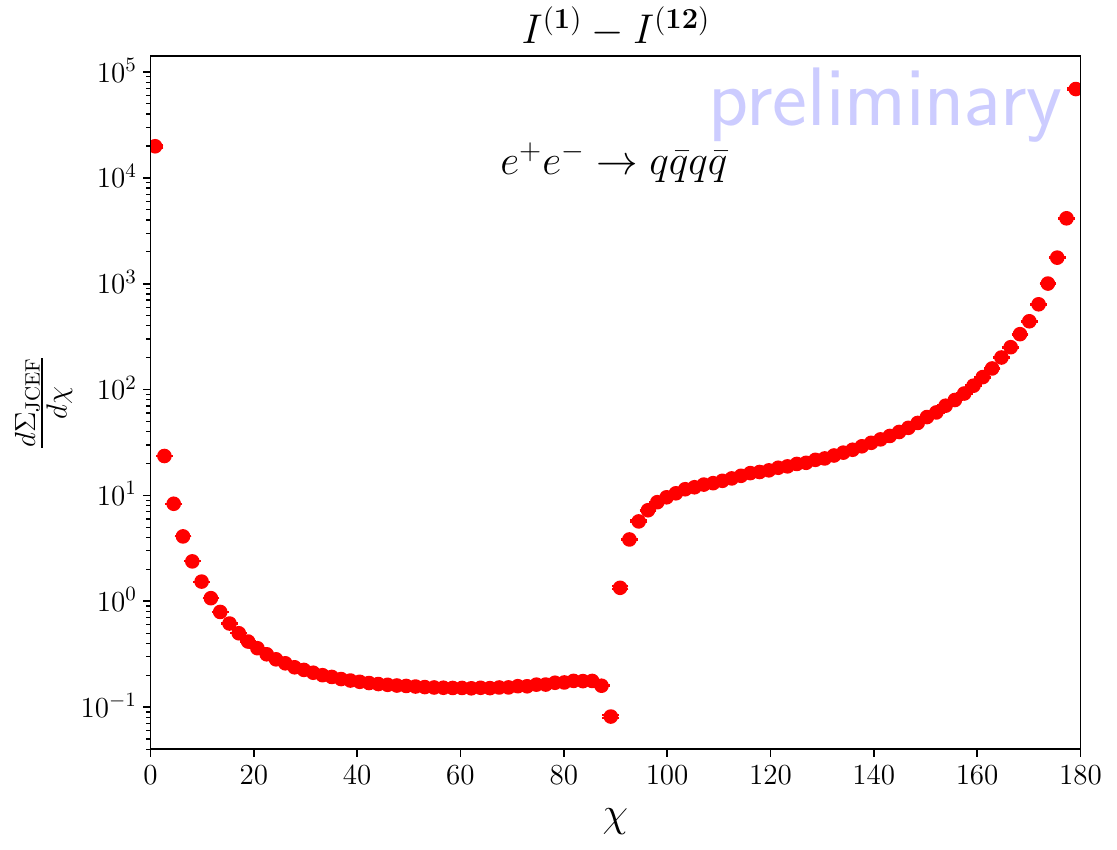}
	\caption{
Real-virtual contributions to the (from left to right)
$\tau$-parameter, $C$-parameter, energy-energy correlation and jet-cone
energy fraction  for $e^+e^-\to q\bar{q}q\bar{q}$ from
$I^{\one}-I^{\otwo}$ terms.
	}
 \label{fig:RV_event_1}
\end{figure}

\begin{figure}[htb]
	\centering
	\includegraphics[width=0.49\textwidth]{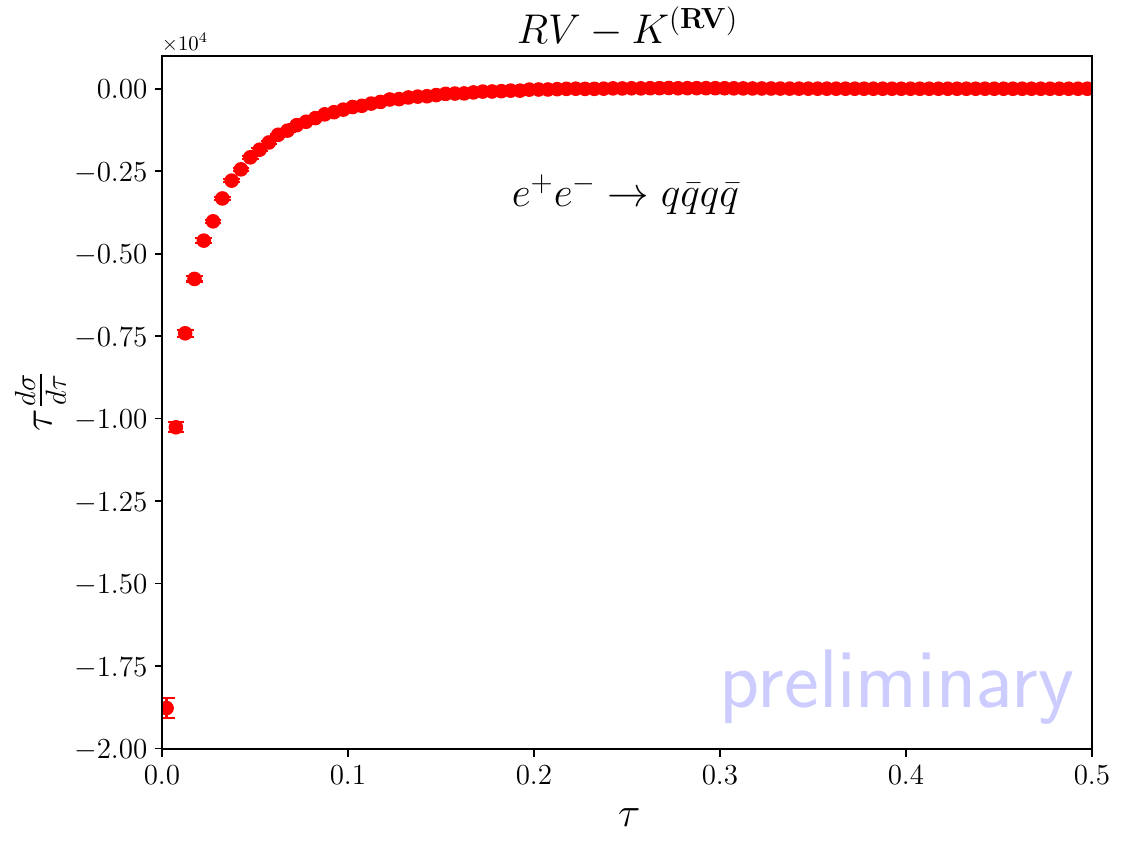}~
	\includegraphics[width=0.49\textwidth]{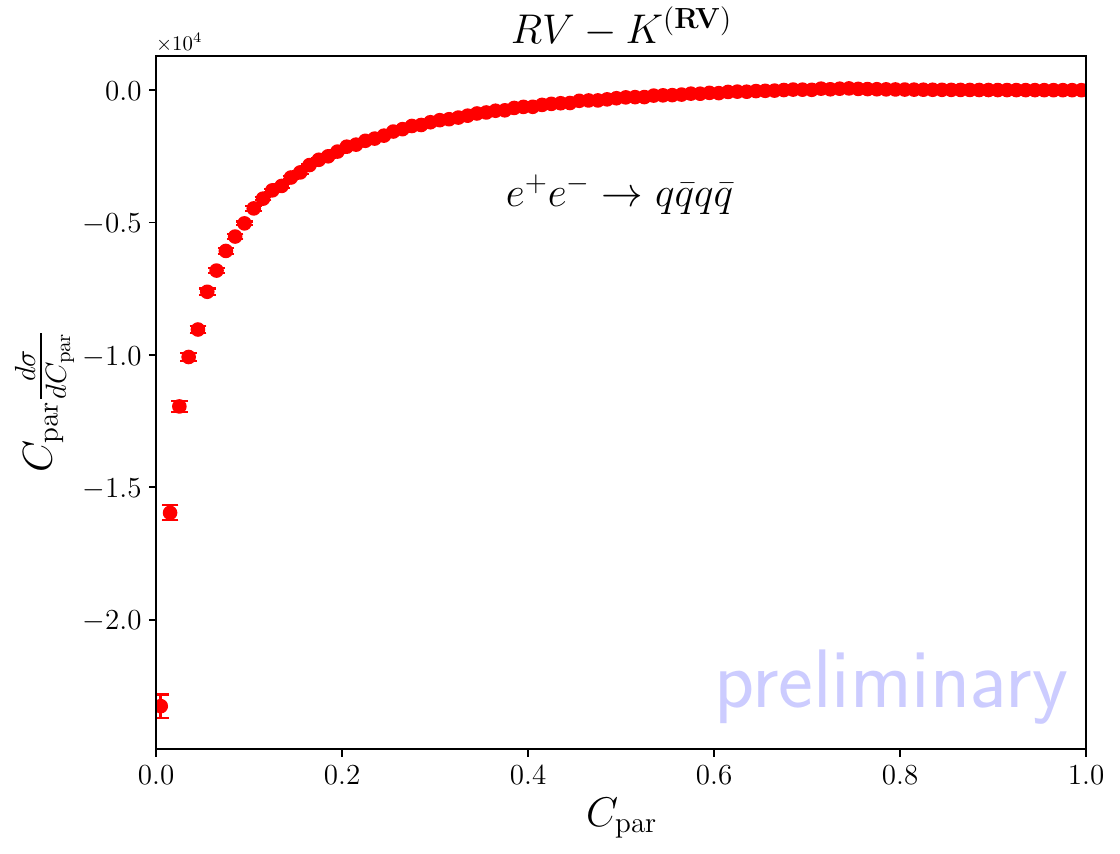}\\
	\includegraphics[width=0.49\textwidth]{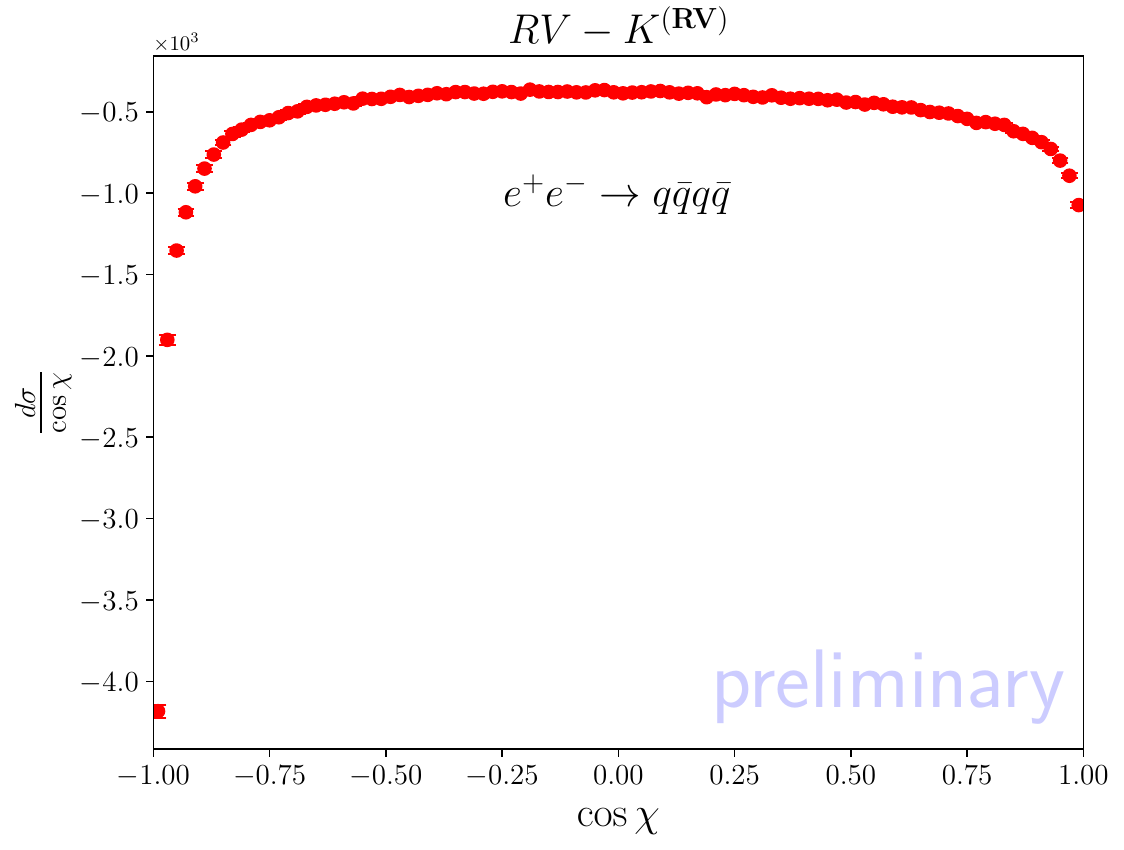}~
	\includegraphics[width=0.49\textwidth]{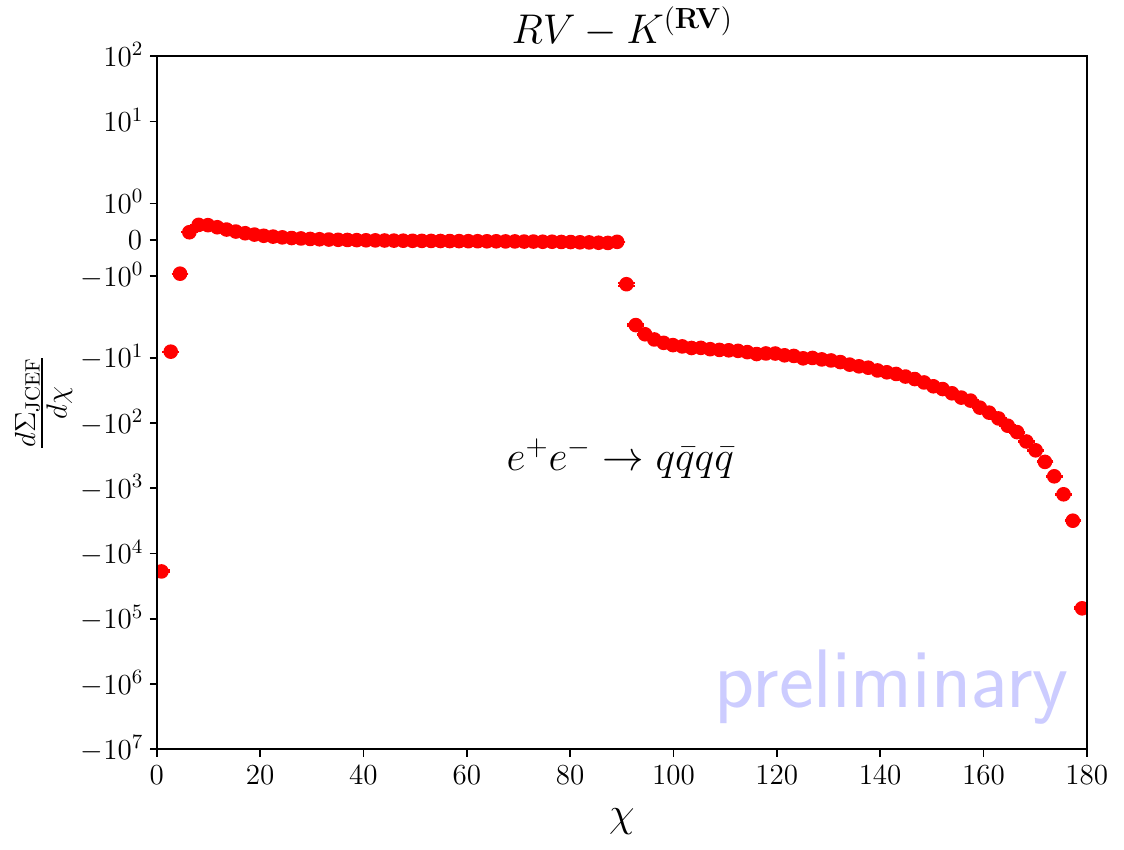}
	\caption{
Real-virtual contributions to the (from left to right)
$\tau$-parameter, $C$-parameter, energy-energy correlation and jet-cone
energy fraction  for $e^+e^-\to q\bar{q}q\bar{q}$ from $RV-K^{\RV}$ terms.
	}
 \label{fig:RV_event_2}
\end{figure}

Next we computed the distributions of the event shape variables
by taking into account the regularized real-virtual contributions.
To perform this study, we utilized the integration grid obtained in the
previous section. To accelerate the speed of the calculation, the integration
was performed $200$ times in parallel with different random number seeds, each
integrating over $10^7$ events. This means that in every bin of each
event shape observable, we have 200 independent estimates. We combine these
estimates using weighted averages described previously, effectively
accumulating the statistics of $2 \times 10^9$ events. The
results are depicted in Figs.~\ref{fig:RV_event_1}~-~\ref{fig:RV_event_2}.
As demonstrated in all cases we observe the good numeric stability.

\section{Conclusions}
We presented a proof-of-concept implementation of the RV and VV contributions
regularized according to the LASS scheme for the NNLO corrections
to $e^+e^- \rightarrow 3$ jets.  Explicit pole cancellation was
demonstrated for the $VV$ contribution.
We have demonstrated that the counterterms correctly approximate
the matrix elements in all IR limits. We computed differential
distributions of event shape variables and found our predictions stable
against variations of the technical cut. This proof-of-concept study
lays the groundwork for developing LASS into a general tool for NNLO
calculations. Future work will focus on optimizing the phase-space
treatment, incorporating initial-state radiation and massive partons,
and automating the set-up.  The LASS scheme is a promising method that
could substantially simplify and speed up precise predictions for
collider observables in the coming years.

\section{Acknowlegdements}
A.K. is supported by the UNKP-21-Bolyai+ New National Excellence
Program of the Ministry for Innovation and Technology from the source
of the National Research, Development and Innovation Fund. A.K. also
kindly acknowledges further financial support from the Bolyai
Fellowship programme of the Hungarian Academy of Sciences.  G.B is
supported by the Hellenic Foundation for Research and Innovation
(H.F.R.I.) under the "2nd Call for H.F.R.I. Research Projects to support
Faculty Members \& Researchers" (Project Number: 02674 HOCTools-II).
\clearpage
\bibliographystyle{JHEP}
\bibliography{LASS_RV_VV}
\end{document}